\documentstyle[aps,prl,twocolumn,floats,epsfig]{revtex}

\providecommand{\TTMF}{\mbox{$\times 10^{-5}$}} 
\providecommand{\KONE}{\mbox{$K^*$}} 
\providecommand{\KONEBAR}{\mbox{$\bar{K}^*$}} 
\providecommand{\KONEZ}{\mbox{$K^{*0}$}} 
\providecommand{\KONEP}{\mbox{$K^{*+}$}} 
\providecommand{\KTWO}{\mbox{$K^*_2$}} 

\providecommand{\YKZG}{\mbox{$88.3{}^{+12.2}_{-11.5}$}} 
\providecommand{\YKPG}{\mbox{$36.7{}^{+8.3}_{-7.6}$}}

\providecommand{\BRKZG}{\mbox{$4.55\ {}^{+0.72}_{-0.68}\ \pm0.34$}} 
\providecommand{\BRKPG}{\mbox{$3.76\ {}^{+0.89}_{-0.83}\ \pm0.28$}} 
\providecommand{\BRKTWOG}{\mbox{$1.66{}^{+0.59}_{-0.53}\pm0.13$}} %

\providecommand{\COMBKGRHOZ}{\mbox{$9.3{}^{+0.6}_{-0.5}$}} 
\providecommand{\COMBKGRHOP}{\mbox{$5.2\pm0.4$}} 
\providecommand{\KSTBKGRHOZ}{\mbox{$5.4\pm0.8$}} 
\providecommand{\KSTBKGRHOP}{\mbox{$2.6\pm0.6$}} 

\providecommand{\ULRRK}{\mbox{0.32}} 

\def\SAKSTP{\mbox{$+0.38{}^{+0.20}_{-0.19}$}}  
\def\BAKSTP{\mbox{$+0.06\pm0.09$}}  

\def\SAKSTZ{\mbox{$-0.13\pm0.17$}} 
\def\BAKSTZ{\mbox{$-0.03\pm0.08$}} 
 
\def\SAKST{\mbox{$+0.08\pm0.13$}} 
\def\BAKST{\mbox{$+0.01\pm0.06$}} 

\begin{document}


\tighten

\title{Study of Exclusive Radiative {\mathversion{bold}$ B$} Meson Decays}

\author{
T.~E.~Coan,$^{1}$ V.~Fadeyev,$^{1}$ Y.~Maravin,$^{1}$
I.~Narsky,$^{1}$ R.~Stroynowski,$^{1}$ J.~Ye,$^{1}$
T.~Wlodek,$^{1}$
M.~Artuso,$^{2}$ R.~Ayad,$^{2}$ C.~Boulahouache,$^{2}$
% ...
}
 
\author{(CLEO Collaboration)}

\address{
$^{1}${Southern Methodist University, Dallas, Texas 75275}\\
$^{2}${Syracuse University, Syracuse, New York 13244}\\
% ...
}

\date{\today}
\maketitle

\begin{abstract} 
We have studied exclusive, radiative $B$ meson decays to 
charmless mesons in $9.7\times 10^6 \
B\bar{B}$ decays accumulated with the CLEO detector. 
We measure 
${\cal B}(B^0\to K^{*0}(892)\gamma) = (\BRKZG)\TTMF$ and 
${\cal B}(B^+\to K^{*+}(892)\gamma) = (\BRKPG)\TTMF$.
We have searched for $CP$ asymmetry in $B\to K^*(892)\gamma$ decays
and measure ${\cal A}_{CP} = \SAKST\pm 0.03$.
We report the first observation of 
$B\to K^*_2(1430)\gamma$ decays 
with a branching fraction of 
$(\BRKTWOG)\TTMF$. 
No evidence for the decays $B\to\rho\gamma$ and
$B^0\to\omega\gamma$ is found and we limit 
${\cal B}(B\to(\rho/\omega)\gamma)/{\cal B}(B\to K^{*}(892)\gamma) < \ULRRK$ 
at 90\% CL.

\end{abstract}


 
The radiative decays, $ B \to K^*(892)\gamma$  and $B\to \rho\gamma$,
occur via the quark transition 
$b \to s,d$ that involves a loop (``penguin'') diagram. In the
Standard Model (SM), the loop amplitude is dominated by a 
virtual intermediate top quark coupling to a $W$ boson and 
probes 
the relative strength of the
$td$ and $ts$ quark couplings ($V_{td}/V_{ts}$)~\cite{ref:ali.braun.simma}. 
The precise determination
of the branching fraction of $ B \to K^*\gamma$~\cite{kstar} can be used to
reduce the theoretical uncertainty in the extraction  of $V_{ub}$ from
the measurement of the decay $B \to \rho \ell \nu$~\cite{isgur,burdman}. The 
magnitudes of the couplings
$|V_{ub}|$ and $|V_{td}/V_{ts}|$ are the lengths of two of the sides of the ``unitarity
triangle'' used to test the SM mechanism of $CP$ violation~\cite{pdg}. 
In addition
the loop amplitude is sensitive to  non-Standard Model (NSM) particles such as
a supersymmetric charged Higgs; the interference of the SM and NSM
amplitudes may result in observable direct $CP$-violating effects manifest
in the charge asymmetry of $B \to K^* \gamma$~\cite{acp.new}. 

The observation of $B\to K^*\gamma$ in 1993
 by the CLEO collaboration~\cite{cleo-btokstgam} 
was the first evidence for $b\to s$ transitions.
The significantly
larger dataset now available allows a more precise determination of this
branching fraction, the first measurement of charge asymmetries in these 
decays and the first search for $B \to \rho\gamma$ and $B^0\to\omega\gamma$
decays. 
In addition we report the first observation of $B\to K^*_2(1430)\gamma$
and the first search for the decay $B^0\to\phi\gamma$ which 
cannot occur 
through a radiative penguin
transition as the decay $B\to\KONE\gamma$. No theoretical
prediction exists in the literature for this decay.

The data were recorded at the Cornell Electron Storage Ring (CESR) with
the CLEO detector~\cite{cleoii,svx}.
The results in this Letter are based upon an integrated luminosity of
9.2~fb$^{-1}$ of $e^+e^-$ data 
corresponding to $9.7\times 10^6$ $B\bar{B}$ meson pairs 
recorded at the $\Upsilon(4{\rm S})$ energy and
4.6~fb$^{-1}$ at 60~MeV below the $\Upsilon(4{\rm S})$ 
energy (``off-$\Upsilon(4{\rm S})$'').
The CLEO detector simulation is based upon GEANT~\cite{geant};
simulated events are processed in the same manner as the
data. The results presented in this Letter supersede the previous
CLEO results~\cite{cleo-btokstgam}.

Candidates for the decays $B\to K_{(2)}^{*}\gamma$ with the subsequent
decays 
$K_{(2)}^{*0}\to K^+\pi^-,\ K^0_{\rm s}\pi^0$ and 
$K_{(2)}^{*+}\to K^+\pi^0,\ K^0_{\rm s}\pi^+$ are selected. 
We define $\KONE$ [$\KTWO$] candidates
by requiring that the $K\pi$ mass be within 110 [120] MeV of
890 [1430] MeV.
We reconstruct the decays $B\to\rho\gamma$ with $\rho^{0,+}\to\pi^+\pi^{-,0}$,
$B^0\to\omega\gamma$ with $\omega\to\pi^+\pi^-\pi^0$ and
$B^0\to\phi\gamma$ with $\phi\to K^+K^-$.
Reference to the charge conjugate states is implicit unless explicitly
stated otherwise. 
The charged-track and $K_{\rm s}^0$ candidates are required to be well
reconstructed and  to originate near the $e^+e^-$ interaction
point (IP). 
Charged kaons and pions are distinguished using the
particle's measured specific ionization ($dE/dx$). We require that the
$dE/dx$ information, when available, is consistent with the appropriate 
hypothesis.
The $K_{\rm s}^0$ candidates are selected through their decay into 
$\pi^+\pi^-$ mesons. The $K_{\rm s}^0$ decay vertex is required to
be displaced from the 
IP, 
and at least one
daughter pion is required to be inconsistent with originating from the
IP. 
Neutral pions are reconstructed from photon pairs
detected in the electromagnetic calorimeter. The photons are required to
have an energy of at least 30 (50)~MeV in the barrel (endcap) region,
and 
the invariant mass of photon pairs
is required to be within three standard
deviations ($\sigma$) of the  $\pi^0$ mass~\cite{pdg}. The high energy
photon from the radiative $B$ decay is required 
to have an energy of at least 1.5~GeV 
and to be in the barrel region $|\cos\theta_\gamma|<0.71$ where
$\theta_\gamma$ is the angle between the beam axis and 
the candidate photon. 

The dominant background comes from continuum ($e^+e^-\to q\bar{q}$ with 
$q = u,c,s,d$) events with high
energy photons originating from initial state radiation
or 
$e^+e^-\to (\pi^0 ,\eta) X$ with $\pi^0,\ \eta\to\gamma\gamma$. 
The $\cos\theta_\gamma$ requirement reduces the former background 
while the latter 
background is suppressed by rejecting
candidate photons that, when combined with an additional
photon candidate, have a mass consistent with the 
$\pi^0$ or $\eta$ mass~\cite{pdg}.
The additional selection criteria described below
reduce backgrounds from non-radiative $B$ decays to a negligible level.
Background from radiative $B$ decays other than the one under study are
discussed later.

We suppress the remaining background from non-radiative $B$ decays 
and continuum by placing requirements on the 
observables $\theta_T$ (the angle between the thrust axis~\cite{thrust} of the 
$B$ candidate and the thrust axis of the
remainder of the event),  
$\theta_B$ (the angle between the $B$ candidate direction and the beam axis), 
$M({\rm R})$ and $\theta_H$
(the mass and helicity angle of the light meson resonance candidate) 
and $dE/dx$. 

Additional background suppression is achieved by requirements on 
the $B$ candidate energy $\Delta E\equiv E({\rm R}) + E(\gamma) - E_{\rm beam} $
and the beam-constrained $B$ mass
$M^2(B)\equiv {E_{\rm beam}^2 - 
({\mathversion{bold}{\bf p}}(\gamma) + 
 {\mathversion{bold}{\bf p}}({\rm R})
)^2}$
where the photon momentum ${\mathversion{bold}{\bf p}}(\gamma)$ is rescaled by
fixing $E(\gamma) = E_{\rm beam} - E({\rm R})$.
The $\Delta E$ [$M(B)$] resolution of 40 MeV [2.8 MeV] is dominated by the 
photon energy resolution [beam energy spread]. 
We select signal and sideband candidates by requiring
$|\Delta E| < 300 \ {\rm MeV}$ and $5.2 < M(B) < 5.3 \ {\rm GeV}$.
If two or more candidates
in an event pass all selection criteria and share daughter tracks or
photons, the candidate with the 
smallest deviation from the nominal resonance mass
is
selected. For the $B\to\rho\gamma$ analysis, the candidate with the 
smallest
$|\cos\theta_B|$ is selected.

We optimize these selection criteria for the $B\to K^*_{(2)}\gamma$
analyses to maximize $S^2/(S+B)$,
where $S$ is the number of
expected signal candidates determined from simulated events 
assuming ${\cal B}(B\to K^*\gamma) = 4.2\TTMF$~\cite{pdg} and 
${\cal B}(B\to K_2^*\gamma) = 1.6\TTMF$~\cite{ref:veseli.olsson}
and $B$ is
the number of background candidates determined from off-$\Upsilon(4{\rm S})$
data. 
For the other analyses the selection criteria are optimized to
yield the smallest upper limit on  
the branching fraction on average using the method 
in Ref.~\cite{feldman-cousins}.

\begin{figure}[htb]
\centering
\epsfig{figure=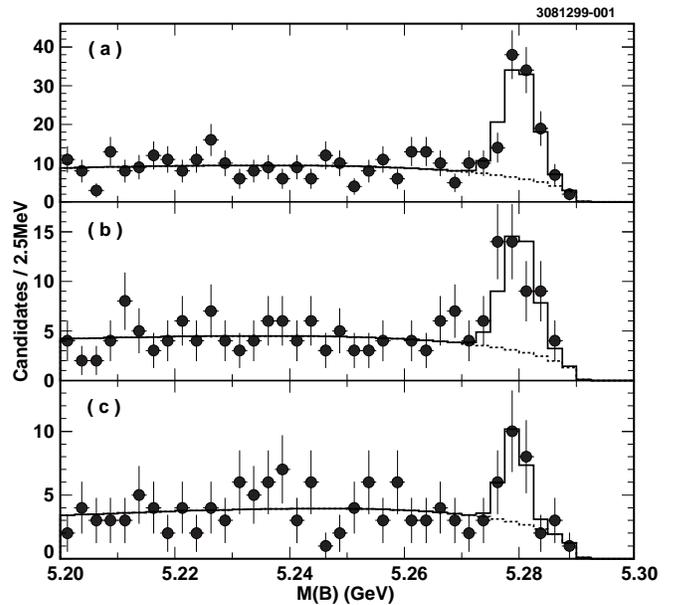,width=\linewidth}
\caption{\label{fig:kst-mb}
Beam-constrained $B$ mass distributions for 
(a) $B^0\to K^{*0}(892)\gamma$,
(b) $B^+\to K^{*+}(892)\gamma$, and
(c) $B \to K^{*}_2(1430)\gamma$.
The data (solid circles) are overlaid with the fit to a
Gaussian and background shape~\protect\cite{argus} (solid line). The
fitted background is indicated by the dashed line.} 
\end{figure}

We perform a simultaneous, binned, maximum-likelihood fit 
to the four $M(B)$ distributions
of 
$B^0\to (K^+\pi^-)\gamma$, 
$B^0\to (K^0_{\rm s}\pi^0)\gamma$,
$B^+\to (K^+\pi^0)\gamma$ and 
$B^+\to (K^0_{\rm s}\pi^+)\gamma$ 
candidates requiring $|\Delta E| < 100 \ {\rm MeV}$.
In the fit the signal component is represented by a Gaussian distribution
and the background is represented by a threshold function~\cite{argus}.
The fitted total yields for $B^0\to \KONEZ\gamma$ and
$B^+\to \KONEP\gamma$ 
are \YKZG\ and \YKPG\ (Fig.~\ref{fig:kst-mb}) and
correspond to branching fractions of 
(\BRKZG)\TTMF\ and (\BRKPG)\TTMF, respectively.
The fractional systematic uncertainties
on the measured branching fractions comprise a common uncertainty of 6.8\%
dominated by the background shape (5\%) and the radiative photon detection
efficiency (3.3\%), and the uncertainties on the reconstruction efficiency
of each $\KONE$ decay mode that range from 2.6\% ($K^0_s\pi^+$) 
to 5.9\% ($K^0_{\rm s}\pi^0$).
The reconstruction efficiency for modes with a charged [neutral]
pion in the final state is 27\% [13\%].
We assume 
${\cal B}(\Upsilon(4{\rm S})\to\bar{B}^0B^0) =  {\cal B}(\Upsilon(4{\rm S})\to B^+B^-)=0.5$
for all branching fractions in this Letter.

 Backgrounds from $B\to {\rm charm}$ are negligible
and 
 backgrounds from charmless two-body  $B$ meson decays are estimated to 
contribute less than 1.2 and 0.6 events to the $ B^0\to \KONEZ\gamma$ 
and $B^+\to \KONEP\gamma$ yields, respectively, based on simulated decays 
and are neglected in the evaluation
of the branching fractions. 
We fit the $M(K\pi)$ distribution summed over \KONEZ\ 
and \KONEP\ within $\pm 150$ MeV of the  $\KONE$ mass~\cite{pdg} to 
search for a nonresonant $B\to K\pi\gamma$ contribution to the 
calculated $B\to \KONE\gamma$ yields. 
No significant nonresonant component with a
threshold shape $\propto\!\! ({M(K\pi)-M(K)-M(\pi)})^{1/2}$
is found, 
but allowing for a nonresonant component
would contribute an additional relative 
uncertainty in the fitted yield of 12\%.
The fitted nonresonant yield is $-16.8\pm 14.7$ events
or less than 23\% of the total yield at 90\% CL.

We search for direct $CP$ violation by measuring the partial rate asymmetry 
${\cal A}_{CP}$,
$$
{\cal A}_{CP} \equiv 
\frac{1}{1-2\eta} 
\, \, \, \, 
\frac
{{\cal Y}(\bar{B}\to\bar{K}^*\gamma)-{\cal Y}(B\to K^*\gamma)}
     {{\cal Y}(\bar{B}\to\bar{K}^*\gamma)+{\cal Y}(B\to K^*\gamma)} \ \ , 
$$
\noindent where ${\cal Y}$ is the fitted yield and $\eta$ is the mistag fraction.
We use the $\KONE$ decay modes $K^+\pi^-$, $K^+\pi^0$ and
$K^0_S\pi^+$ to measure 
${\cal A}_{CP}$. 
In these decay modes the
charge of the kaon or the $K^*$ contains unambiguous information 
about  the
$B$ flavor. 
Only the $K^+\pi^-$ decay mode has a mistag rate significantly different
from zero as determined from simulated events.
Mistagging in this mode is due to the 100\% transverse
polarization of the \KONEZ\ from $B^0\to \KONEZ\gamma$ decays
that results in a $\sin^2\theta_{\rm H}$ distribution.
This distribution favors nearly equal momenta of $\sim\!\! 1.2 \ {\rm GeV}/c$
for the charged kaon and pion from the  {\KONE}.
The kaon and pion cannot be kinematically distinguished
when  $p_K \approx p_\pi$, and their expected $dE/dx$ is nearly 
identical in this momentum range.
We exclude these ambiguous $\KONEZ$ 
candidates from the ${\cal A}_{CP}$ measurement
by requiring $|p(K)-p(\pi)| >0.5 \ {\rm GeV}/c$. 
This requirement 
minimizes the 
statistical uncertainty 
on ${\cal A}_{CP}$ in the 
$K^+\pi^-$ decay mode with $\eta = (3.45\pm0.02)\%$ 
and a relative efficiency of $(62.0\pm0.5)\%$ as 
determined from simulated events.

To measure ${\cal A}_{CP}$, 
we fit 
the $M(B)$ distributions of $B\to \KONE\gamma$ and $\bar{B} \to
\KONEBAR\gamma$ 
candidates simultaneously for both neutral and charged $B$ meson
decays to extract the total yield and asymmetry of
both the $B \to \KONE\gamma$ signal and the background in the range
\mbox{$5.2 < M(B) < 5.3 \ {\rm GeV}$}
with a procedure similar to that described for the $B\to\KONE\gamma$
branching fractions.
For neutral and charged $B\to \KONE\gamma$
decays, we determine ${\cal A}_{CP} = \SAKSTZ$ 
and \SAKSTP, respectively, for the
signal and \BAKSTZ\ and \BAKSTP\ for the background.
The asymmetry for the sum of neutral and charged $B\to \KONE\gamma$
decays is \SAKST\ [{\BAKST}] for the signal [background].
Systematic searches for detector- or reconstruction-induced
charge asymmetries for charged pions and kaons revealed no significant
bias ($|\Delta{\cal A}_{CP}| < 1.5\%$). In addition, studies of simulated 
$B\to \KONE\gamma$ decays indicate that cross-feed between
different $\KONE$ decay modes is $<1\%$. 
Our conservative estimate of the 
systematic uncertainty on ${\cal A}_{CP}$ is 2.5\%.

 Radiative $B$ meson decays to the \KTWO\ and the nearby 
$K^*(1410)$  can be distinguished by the helicity angle distributions
($\propto\! \cos^2\theta_{\rm H} - \cos^4\theta_{\rm H}$ and
 $\propto\! \sin^2\theta_{\rm H}$, respectively) as well as the
resonance widths of $\sim\!\!100$ and 
$\sim\!\! 230\ {\rm MeV}$~\cite{pdg}. 
We fit the $M(B)$ distributions
of candidates that pass [fail]
the requirement  $|\cos\theta_{\rm H}| < {\cal H}$
designed to enhance [deplete] 
$B\to\KTWO\gamma$ decays
where ${\cal H}$ ranges from 0.20 to 0.30
depending on the \KTWO\ decay mode.
The overall efficiency for 
passing [failing] the helicity angle requirements
is $(10.1\pm0.3)\%$ [$(1.09\pm0.08)\%$] and
$(0.80\pm0.13)\%$ [$(0.59\pm0.10)\%$] for simulated 
$B\to \KTWO\gamma$ and $B\to K^*(1410)\gamma$ decays, respectively,
where the quoted efficiency includes 
${\cal B}(\KTWO\to K\pi)=(49.9\pm 1.2)\%$
and ${\cal B}(K^*(1410)\to K\pi)=(6.6\pm 1.3)\%$~\cite{pdg}.
The simultaneous determination of  ${\cal B}(B\to \KTWO\gamma)$
and  ${\cal B}(B\to K^*(1410)\gamma)$ from the two fitted yields and 
the quoted efficiencies shows that 
  ${\cal B}(B\to \KTWO\gamma)$ is significant at over 
$3 \sigma$
for the most probable value of  
${\cal B}(B\to K^*(1410)\gamma)$ whilst 
${\cal B}(B\to K^*(1410)\gamma)$ is 
less than 
$1 \sigma$
significant for the most probable value of  
${\cal B}(B\to \KTWO\gamma)$. We therefore interpret the
signal as being due to $B\to \KTWO\gamma$ only and 
determine
${\cal B}(B\to K^*(1410)\gamma) < 12.7 \TTMF$
at 90\% CL.  
The $M(B)$ distribution of $B\to \KTWO\gamma$ candidates 
passing the $|\cos\theta_{\rm H}|$ requirements is shown
in Fig.~\ref{fig:kst-mb}(c) summed over the charged and neutral 
$\KTWO$ meson decays. 
The fitted yield of $15.9{}^{+5.7}_{-5.1}$ 
events is significant at $4.3\sigma$ [$3.3\sigma$]
before [after] 
inclusion of systematic uncertainties.
Assuming equal decay rates to  charged and neutral $\KTWO$, 
the yield corresponds to a branching fraction of 
(\BRKTWOG)\TTMF, where the systematic uncertainties are evaluated
as described for the $B\to \KONE\gamma$ branching fractions.

 The branching fractions of $B\to \KONE\gamma$ and 
$B\to \KTWO\gamma$ have been predicted by two 
groups~\cite{ref:veseli.olsson,ref:ali.ohl.mannel}
and differ in the treatment of long distance effects 
on the form factors. 
The minimal uncertainty is achieved by the ratio 
${\cal B}(B\to \KTWO\gamma)/{\cal B}(B\to \KONE\gamma) =  
0.39{}^{+0.15}_{-0.13}$  
that compares favorably with the prediction of 
Veseli and Olsson of  $0.37\pm 0.10$~\cite{ref:veseli.olsson,ref:olsson}
and disagrees with the Ali, Mannel, and Ohl 
range of 3.0 to 4.9~\cite{ref:ali.ohl.mannel}.

\begin{figure}[htb]
\centering
\epsfig{figure=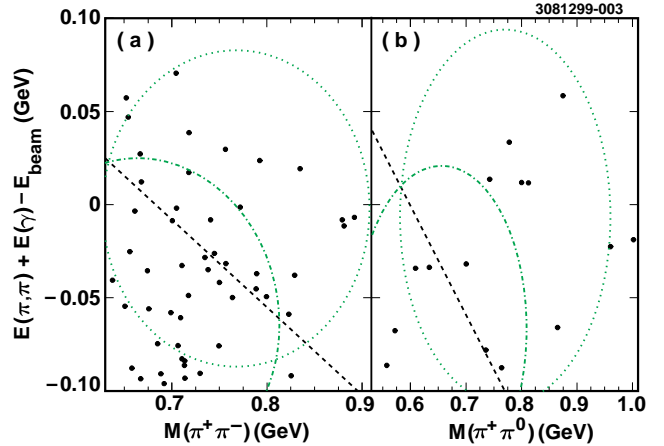,width=\linewidth}
\caption{\label{fig:rho}
  The $\Delta E$ {\it vs} $M(\pi\pi)$ distributions for a) $B^0\to\rho^0 \gamma$
and b) $B^+\to \rho^+\gamma$ candidates. Candidates above the diagonal dashed
line survive the final selection criterion.
The dotted [dot-dash] line approximates the limits that would
contain 90\% of the $B\to\rho\gamma$ [$B\to K^*\gamma$] candidates.
}
\end{figure}

In order to limit $|V_{td}/V_{ts}|$, we have searched for the
decays $B\to \rho\gamma$ and $B^0\to \omega\gamma$. The 
 $\rho\gamma$ final states suffer from background both from continuum 
and from $B\to \KONE\gamma$ when a charged kaon is
misidentified as a pion. 
Continuum is the only significant background to
$B\to\omega\gamma$. 
The $\Delta E$ {\it vs.} $M(\pi\pi)$ distributions
for $B^0\to\rho^0\gamma$ and $B^+\to\rho^+\gamma$ candidates are shown
in Fig.~\ref{fig:rho} after a requirement of $5274 < M(B) < 5286 \ {\rm MeV}$.
The $\KONE$ background peaks in the lower left hand corner of each
distribution whilst the signal peaks near the center and the continuum
background is constant. Twenty-four [ten] candidates survive 
 the requirement of $\Delta E > -0.47 M(\pi\pi) + 0.32\ {\rm GeV}$ 
[$\Delta E > -0.58 M(\pi\pi) + 0.35\ {\rm GeV}$]
for $B^0\to\rho^0\gamma$ [$B^+\to\rho^+\gamma$]
as shown in Fig.~\ref{fig:rho}. We estimate the combinatorial
background from  fits to the $M(B)$ distributions
and the background from  $B\to \KONE\gamma$ 
using the measured branching fractions
and the reconstruction efficiency from simulated 
$B\to \KONE\gamma$  decays. The overall reconstruction efficiency is 
$(12.8\pm0.7)\%$ [$(8.5\pm0.6)\%$] and the background comprises
 \COMBKGRHOZ\ [\COMBKGRHOP] continuum events and  
\KSTBKGRHOZ\ [\KSTBKGRHOP]  $B\to \KONE\gamma$  events
for the $\rho^0$ [$\rho^+$] decay mode. 
We determine  upper limits 
of ${\cal B}(B^0\to\rho^0\gamma) < 1.7\TTMF$ 
and ${\cal B}(B^+\to\rho^+\gamma) < 1.3\TTMF$ at 90\% CL.
All branching fraction upper limits in this Letter are determined with 
the method in \cite{feldman-cousins} after 
reducing the central values of the 
estimated background, efficiency, daughter branching
fractions and  number of $B\bar{B}$ pairs by one 
standard deviation.

\begin{figure}[htb]
\centering
\epsfig{figure=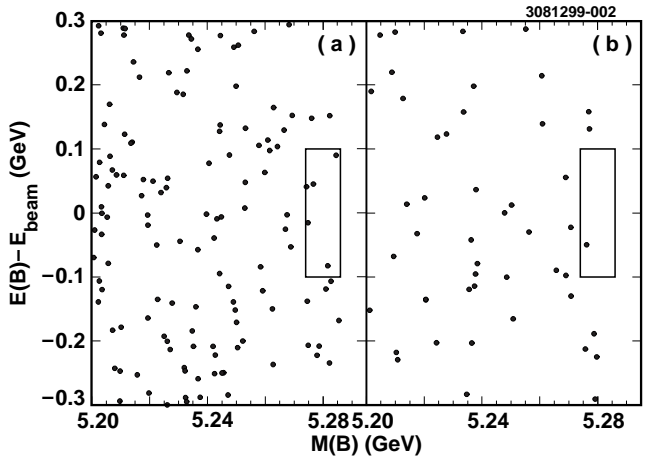,width=\linewidth}
\caption{\label{fig:omega.phi} The $\Delta E$ {\it vs} beam-constrained
$B$ mass distributions for (a) $B^0\to\omega\gamma$ and (b) $B^0\to\phi\gamma$
candidates. The rectangular area indicates the signal region.
}
\end{figure}

We observe 5 $B^0\to\omega\gamma$ candidates in the signal
region $|\Delta E|< 100 \ {\rm MeV}$ and 
 $5274 < M(B) < 5286 \ {\rm MeV}$ shown in Fig.~\ref{fig:omega.phi}(a).
The  combinatorial background is estimated to be  $2.68{}^{+0.13}_{-0.12}$
from the fit to the $M(B)$ distribution. This corresponds to 
${\cal B}(B^0\to\omega\gamma) < 0.92\TTMF$ at 90\% CL
with the reconstruction efficiency of $(9.7\pm0.8)\%$.

We determine an upper limit on ratio 
$R \equiv {{\cal B}(B\to\rho\gamma)}/{{\cal B}(B\to \KONE\gamma)}$
from the likelihood ${\cal L}(R)$ where
${\cal B}(B\to\rho\gamma) \equiv {\cal B}(B^+\to\rho^+\gamma)  
 = 2{\cal B}(B^0\to\rho^0\gamma) = 2{\cal B}(B^0\to\omega\gamma)$ 
and ${\cal B}(B\to \KONE\gamma)$ is the average over $B^+$ and 
$B^0$ decays. 
The 90\% CL limit on $R$, $R_{90}$, is given 
by $\int_0^{R_{90}} {\cal L}(R) dR /\int_0^\infty {\cal L}(R) dR  = 0.90$
where ${\cal L}(R) = \prod_i e^{-\mu_i}\mu_i^{n_i}/{n_i!} $ with
$i = \rho^+,\rho^0,\omega$; $n_i =$ total number of $B\to\rho\gamma$
candidates and $\mu_i = b_i^{\rm c} + b_i^{\rm K} + 
N(B\bar{B})\, \epsilon_i \, {\cal B}_i^s\, R\, {\cal B}(B\to \KONE\gamma)$.
The estimated continuum [$B\to \KONE\gamma$] background
is $b_i^{\rm c}$  [$b_i^{\rm K}$], $\epsilon_i$ is
the reconstruction efficiency and ${\cal B}_i^s$ is
the daughter branching fraction.
Similarly, we form ${\cal L}(|V_{td}/V_{ts}|)$ using
the relationship $|V_{td}/V_{ts}|^2 = {R/\xi}$
where  $\xi$ is the ratio of the $B\to\rho\gamma$ 
and $B\to \KONE\gamma$ form factors.
The upper limit of $R < \ULRRK\ $(0.36)  
corresponds to 
$|V_{td}/V_{ts}| < 0.72\ $(0.76) at 90 (95)\% CL 
for $\xi = 0.58$~\cite{ref:ali.braun.simma}.  
Other estimates of $\xi$
are 0.77~\cite{ref:narison} and $0.81\pm0.09$~\cite{ref:soares}.
Our evaluation of a $|V_{td}/V_{ts}|$ limit assumes that these decays
proceed via top-quark-dominated electromagnetic penguin transitions
and neglects possible contributions from 
final state interactions~\cite{ref:donoghue.golowich.petrov},
$W$-exchange~\cite{ref:cheng.and.eilam.ioan.mendel} 
or $W$-annihilation~\cite{ref:ali.braun}.

We observe one $B^0\to\phi\gamma$ candidate in the signal
region $|\Delta E|< 100 \ {\rm MeV}$ and 
 $5274 < M(B) < 5286 \ {\rm MeV}$ shown in Fig.~\ref{fig:omega.phi}(b).
We estimate the combinatorial background to be $1.2\pm0.1$
events from the fit to the $M(B)$ distribution.
This corresponds to ${\cal B}(B^0\to\phi\gamma) < 0.33\TTMF$ at 90\% CL
with the reconstruction efficiency of $(23.0\pm 0.6)\%$.

In summary, the $B\to K^*(892)\gamma$ branching fractions
have been measured with improved precision. 
A new radiative decay mode $B\to K^*_2(1430)\gamma$ 
has been observed and found to agree with one of two
theoretical predictions. The partial rate asymmetries 
in $B\to K^*(892)\gamma$ decays are measured with a precision
of better than 20\% and found  to be consistent
with Standard Model expectations. We find no evidence 
for the process $b\to d\gamma$ and determine a limit on the ratio of 
${\cal B}(B\to\rho\gamma)/{\cal B}(B\to K^*(892)\gamma) < \ULRRK$
at 90\% CL. Using a model-dependent derivation of
the ratio of the $B\to\rho\gamma$ and $B\to K^*(892)\gamma$ form factors, 
the ratio of branching fractions implies that 
$|V_{td}/V_{ts}|< 0.72$ at 90\% CL.

We thank A.~Ali, T.~Mannel, M.~Neubert, M.G.~Olsson and S.~Veseli
for useful discussions.
We gratefully acknowledge the effort of the CESR staff in providing us with
excellent luminosity and running conditions.
This work was supported by 
the National Science Foundation,
the U.S. Department of Energy,
the Research Corporation,
the Natural Sciences and Engineering Research Council of Canada, 
the A.P. Sloan Foundation, 
the Swiss National Science Foundation, 
and the Alexander von Humboldt Stiftung.


\begin{thebibliography}{99}

\bibitem{ref:ali.braun.simma}
A.~Ali, V.M.~Braun, and H.~Simma,
Z. Phys. C {\bf 63}, 437 (1994).

\bibitem{kstar}
We refer to $K^*(892)$ as \KONE\ and $K^*_2(1430)$ as {\KTWO}.

\bibitem{isgur}
N.~Isgur and M.B.~Wise,
Phys. Rev. D {\bf 42}, 2388 (1990). \relax

\bibitem{burdman}  
G.~Burdman and J.F.~Donoghue,
Phys. Lett. B {\bf 270}, 55 (1991). \relax

\bibitem{pdg} 
Particle Data Group, C.~Caso {\em et al.}, Eur.~Phys.~J. C {\bf  3}, 1 (1998) 
      and 1999 off-year partial update for the 2000 edition available on 
      the PDG WWW pages (URL: {\tt http://pdg.lbl.gov/}). 
\relax

\bibitem{acp.new}
L.~Wolfenstein and Y.L.~Wu, 
Phys. Rev. Lett. {\bf 73}, 2809 (1994);
H.M.~Asatrian and A.N.~Ioannissian,
Phys. Rev. D {\bf 54}, 5642 (1996);
A.L.~Kagan and M.~Neubert,
Phys. Rev. D {\bf 58}, 094012. 


\bibitem{cleo-btokstgam}
R.~Ammar {\em et al.}, 
Phys. Rev. Lett. {\bf 71}, 674 (1993). \relax 

\bibitem{cleoii} 
Y.~Kubota {\em et al.}, 
Nucl. Instrum. Methods A {\bf 320}, 66 (1992). \relax

\bibitem{svx}
T.~Hill, Nucl. Instrum. Methods A {\bf 418}, 32 (1998). \relax

\bibitem{geant}
R.~Brun {\em et al.}, {\sc GEANT} 3.15, CERN Report No.~DD/EE/84-1 (1987). \relax

\bibitem{thrust}
E.~Farhi, Phys. Rev. Lett. {\bf 39}, 1587 (1977). \relax

\bibitem{ref:veseli.olsson}
S.~Veseli and M.G.~Olsson,
Phys. Lett. B {\bf 367}, 309 (1996).

\bibitem{feldman-cousins}
        G.~Feldman and R.~Cousins,
        Phys. Rev. D {\bf 57}, 3873 (1998). \relax


\bibitem{argus}
$f(x) \propto  x \sqrt{1-x^2} \exp (\kappa (1-x^2)) $ where 
$x \equiv M(B)/E_{\rm beam}$. The parameter 
$\kappa$ is determined by the fit.
H.~Albrecht {\it et al.}, 
Phys. Lett. B {\bf 241}, 278 (1990); {\bf 254}, 288 (1991).


\bibitem{ref:ali.ohl.mannel}
A.~Ali, T.~Ohl, and T.~Mannel,
Phys. Lett. B {\bf 298}, 195 (1993).

\bibitem{ref:olsson}
The uncertainty on the ratio of branching fractions
is dominated by the additional fractional uncertainty
in ${\cal B}(B\to K^*_2\gamma)$, M.~G.~Olsson (private communication).

\bibitem{ref:narison}
S.~Narison, Phys. Lett B {\bf 327}, 354 (1994).

\bibitem{ref:soares}
J.~M.~Soares, Phys. Rev. D {\bf 49}, 283 (1994).

\bibitem{ref:donoghue.golowich.petrov}
J.F.~Donoghue, E.~Golowich and A.A.~Petrov,
Phys. Rev. D {\bf 55}, 2657 (1997).

\bibitem{ref:cheng.and.eilam.ioan.mendel} 
H.-Y.~Cheng, 
Phys. Rev. D {\bf 51}, 6228 (1995);
G.~Eilam, A.~Ioannissian and R.R.~Mendel,
Z. Phys. C {\bf 71}, 95 (1996).

\bibitem{ref:ali.braun}
A.~Ali and V.M.~Braun,
Phys. Lett. B {\bf 359}, 223 (1995).

\end{thebibliography}
\end{document}